\documentclass[rsi,preprint,showpacs,showkeys]{revtex4-1}
\usepackage{graphicx}
\usepackage[dvips]{pstricks}
\usepackage{bm}
\usepackage[USenglish]{babel}
\usepackage{amsmath}
\begin{document}
\title[Glucose monitoring in human subjects using PSOCT]{Polarization Sensitive Optical Coherence Tomography for Blood Glucose Monitoring in Human Subjects} 
\author{Jitendra Solanki} 
\address{Department of Applied Physics, \\ 
Shri G S Institute of Technology \& Science, \\ Indore - 452 003 India.}
\author{Om Prakash Choudhary} 
\address{Department of Applied Physics, \\ 
Shri G S Institute of Technology \& Science, \\ Indore - 452 003 India.}

\author{P. Sen}
\address{Laser Bhawan, School of Physics,\\ 
Devi Ahilya University, Khandwa Road, \\  Indore - 452 007 India.}

\author{J. T. Andrews}
\email{jtandrews@sgsits.ac.in}
\address{Department of Applied Physics,\\ 
Shri G  S Institute of Technology \& Science, \\ Indore - 452 003 India.}

\date{Received: date / Accepted:date}

\begin{abstract} 
A device based on Polarization sensitive optical coherence tomography is developed to monitor blood glucose levels in human subjects. The device was initially tested with tissue phantom. The measurements with human subjects for various glucose concentration levels are found to be linearly dependent on the degree of circular polarization obtainable from the PS-OCT.\end{abstract}


\maketitle
\section{Introduction}

The optically active nature of glucose inspires us to use polarization state of light as key for blood glucose monitoring. In an optically active medium the velocities of right circularly polarized (RCP) and left circularly polarized (LCP) lights differ. Consequently, such medium has the ability to rotate the plane of polarization of the resultant linearly polarized (LP) light. Larger glucose concentration is expected to yield larger polarization rotation. Despite the wealth of different properties that can be probed with polarized light, the turbid nature of  biological samples makes them nontrivial. For example multiple light scattering in turbid media can hinder measurement and data interpretation. Over the years, many attempts were made to overcome these complexities involved in the measurement of polarization properties of scattered light from biological tissues.

Several techniques are proposed and adopted to interpret the data obtained from turbid media. According to Ghosh et al. \cite{Ghosh}, the study of polarization properties of biological tissue using light scattering carries a wealth of morphological and functional information and has  potential biomedical importance. They discussed comprehensive turbid polarimetry platform consisting of forward Monte Carlo modeling and inverse polarization decomposition analysis.
Hee et al. \cite{Hee}, used optical coherence domain reflectometer for characterizing phase retardation between orthogonal linear polarization modes at each reflection point in a birefringent sample. They used the reflectometer to characterize attenuation and birefringence of tunic media in calf coronary artery. Kohl et al. \cite{Kohl}, showed that presence of glucose dissolved in an aqueous solution increases refractive index of the solution and also influences scattering properties of any particles suspended within it. 
They demonstrated the effect of the glucose on light transport in highly scattering tissue simulating phantoms using diffusion theory. Their results pave way to use this effect for noninvasive glucose monitoring in diabetic patients. The next step in this direction was to correlate scattering coefficient with blood glucose concentration. In 1997 Bruulsema et al. \cite{Bruu}, examined reduced scattering coefficient of tissue in response to steep changes in blood glucose levels of diabetic volunteers. They observed correlation between steep changes in blood glucose concentration and tissue reduced scattering coefficients in 30 out of 41 subjects. Wang et al. \cite{Wang} reported existence of linear relationship between angles of rotation in Muller matrix elements and concentration of glucose. 

In a pilot study of specificity of noninvasive blood glucose monitoring using Optical Coherence Tomography (OCT) technique, Larin et al. \cite{Larin} report that several osmolytes may change refractive index mismatch between interstitial fluid and scattering centers in tissue. They found that the effect of glucose is approximately two to three orders higher and concluded that OCT technique may provide blood glucose monitoring with sufficient specificity under normal physiological conditions. They also used phase-sensitive optical low-coherence reflectometry (PS-OLCR) for measurement of analyte concentration \cite{PSOCT} and demonstrated a high degree of sensitivity and accuracy of phase measurements of analyte concentrations in their experiment. Cameron and Li \cite{Came} examined the polarization sensitivity of glucose concentration in physiological range for highly scattering biological media and concluded that OCT based glucose monitoring is approaching standard invasive and minimally invasive techniques. Lee et al. \cite{Lee} measured changes of degree of circular polarization (DOCP) in intralipid suspensions and human cervical tissue using polarization sensitive OCT (PS-OCT) system and proposed that this technique may provide a unique diagnostic tool.
      
Motivated by the studies reported above, we have used PS-OCT technique for analyzing blood glucose concentration in tissue phantom and noninvasive blood glucose in human subjects. We report the measurement of change in degree of polarization as a function of glucose concentration in tissue phantom as well as in human subjects. The experimental observations have been explained using Mie scattering theory by incorporating the refractive index dependence on the state of polarization in the optically active medium.

\section{Experiment} 

\begin{figure}[htb]
	\begin{center}
		\includegraphics[width=.8\columnwidth]{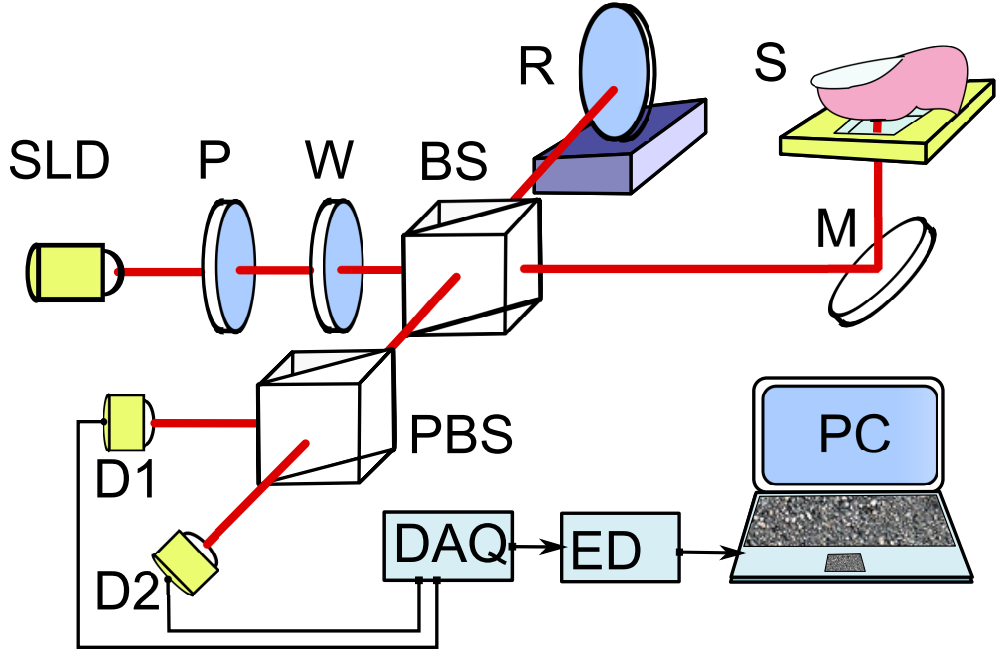}
	\end{center}
	\caption{Schematic of the Polarization Sensitive Optical Coherence Tomography setup: SLD~-~Superluminescent diode, P-Polarizer, W-Quarter Wave plate, BS-Non-polarizing Beam Splitter, S-Sample Arm, R-Reference arm with scanning assembly, PBS-Polarizing Beam Splitter, D1,D2-Photo Diodes, DAQ-Data acquisition system, ED-Envelope detector, PC-Laptop and other electronic processing units.}
\end{figure}
  
The optical coherence tomography setup developed by us is custom designed
for the measurements with tissue phantom, blood samples as well as for non-invasive measurements with human subjects. A schematic of the Michelson interferometer type PS-OCT is shown in Figure 1. A low coherence broad band light source viz; Hamamatsu Superluminescent diode (SLD) having a center wavelength of 835 nm and bandwidth 50 nm is used. The sample arm of
the device has the option to accommodate samples of tissue phantom or blood
or any one finger of the subject for non-invasive measurements.  Low coherence beam of light emitted from SLD was passed through the linear polarizer and a quarter wave plate at 45$^{\circ}$ to the polarizer axis as shown in the figure to provide circularly polarized (CP) light. The CP beam was delivered to the a 50-50 non-polarizing beam splitter (BS). The reference arm was scanned repetitively at a constant speed of (3 mm/s). The back reflected light from the reference arm and the backscattered light from the sample arm form an interferogram, which is delivered to the polarizing beam splitter (PBS). The orthogonal components of the interference signal are detected by two photo diodes $D_1$ and $D_2$. These orthogonal components of the OCT signal were filtered, amplified and delivered to a laptop for further processing and analysis. 

\subsection{Signal Processing}

The sample arm of the device consists of a reference glass plate. As shown in Figure 1, the sample under study makes direct contact with the top surface of the glass plate. The OCT signal obtained from samples such as tissue phantom or human subject is expected to give two distinct peaks, corresponding to the air/glass and glass/tissue interfaces. The signal obtained for air/glass interface is a pure Gaussian having signatures of the light source while the signal obtained for glass/tissue interface have the signatures of the sample under study. Since the optically active medium rotates the LCP and RCP at different magnitudes, finite degree of circular polarization occurs for different values of glucose concentrations. Accordingly, the PS-OCT setup employs a polarizing beam splitter (PBS) to disentangle the orthogonal components of the light polarization. The amplitudes  $E_x [= (\sigma_{+}+\sigma_{-})/2]$ and $E_y [= (\sigma_{+}-\sigma_{-})/2]$ of the backscattered signal from the sample are measured using the photo diodes $D_1$ and $D_2$ with $\sigma_{+(-)}$ being the electric filed component of the RCP (LCP). 
 
\subsection{Tissue Phantom}  

Intralipid is widely used in many optical experiments to find the scattering properties of biological tissues and can be used as a good scatterer like RBCs in human blood. 
Intralipid is an emulsion of soybean oil, egg phospholipids and glycerin. The major advantages of intralipid are its well known optical properties and the similarity of its microparticles to lipid cell membranes and organelles that constitute the source of scattering in biological tissue. We used intralipid as a tissue phantom that provides the backscattered component of an incident light. Average size of scatterers in intralipid measured using laboratory confocal microscope is found to be 3.5 $\mu$m with refractive index of 1.42.

The measurement with tissue phantom was taken under the hypoglycemic, normal, and hyperglycemic glucose levels of humans as ($<$ 80 mg/dl), (80-110 mg/dl), and ($>$110 mg/dl), respectively. Following measurement modality is adopted during the experiment for use with tissue phantom.  First, a fixed volume (1 ml) of intralipid (20\% w/V) was diluted to 100 ml using distilled water and the diluted solution of 0.01\% of tissue phantom was prepared. In the second step 200 mg of glucose was dissolved in 100 ml distilled water to make glucose solution. In the third step 100$\mu$l of glucose solution was added to 1 ml of tissue phantom using U-TEK Chromatography syringe with least count of 5$\mu$l. This gives us 20mg/dl glucose concentration. Two minutes of settlement time was given between tissue phantom and every glucose concentrations. The glucose concentration was increased upto 200mg/dl. Only 10$\mu$l of whole solution was placed on 140$\mu$m thick coverslip in the sample arm and five measurements of backscattered light were taken for each concentration.

The results obtained from the intralipid were verified using Mie scattering 
theory. We have further extended the analysis with voluntary human subjects.   

\subsection{Blood samples from human subjects}

Our device has the flexibility of monitoring blood glucose invasively as well as non-invasively. In the present case, the thumb finger of the subject is used in the sample arm instead of tissue phantom in the previous case. Before placing the finger the distal pulp region of thumb was cleaned by mild spirit and glycerin was applied for better optical contact between thumb skin of human subject and coverslip. Measurements were performed simultaneously using PS-OCT setup and with standard glucometer or through chemical route (GOD/BOD Method).

It is well known that after a meal, a temporary rise in blood sugar occurs. The extent and duration of sugar level depends on the characteristics of the food as well as the subjects. For a normal subject the increase in blood sugar level may not exceed 160-180mg/dl. In case of diabetic subject the rise and reduction of glucose level after meal is slower and elevated. The subjects volunteered in the current experiment are asked to consume 100g glucose with water. After a settling time of 10 minutes, the measurement of blood glucose were carried out using  PSOCT system and through glucometer, simultaneously. The measurement were repeated for every 10-15 minute interval. A correlation between the glucose concentration and DOCP is obtained and analyzed. 

\section{Theoretical Analysis}

A theoretical model is developed to understand the experimental observations. Intralipid and human blood can be theoretically categorized as turbid media in which the backscattered light from sample arm contributes to OCT signal. The signal from the sample arm comprises of single and multiple scattered components of light. Strong multiple scattering typical of most biological tissues leads to loss of phase, direction and polarization of incident radiation. It has been observed that for circularly polarized light randomization of polarization requires more scattering events than the randomization of its direction \cite{Cai,Ni,Xu}. Multiply scattered polarized light from a turbid medium carries information about refractive index variation. The incident light in the present experimental setup is circularly polarized. Tinoco et al. \cite{Tin} discussed differential scattering of circularly polarized (CP) light and concluded that circular differential scattering retains information concerning the chiral properties of the scattering object even when the sample is partially or completely disordered. The circular intensity differential scattering (CIDS) or degree of circular polarization (DOCP) is defined as \cite{Born,jsol} 
\begin{equation}
DOCP = \frac{I_L-I_R}{I_L+I_R}=\frac{\sigma_L-\sigma_R}{\sigma_L+\sigma_R}=\frac{I_x}{I_y}
\end{equation}      
Here, $\sigma_L$ and $\sigma_R$ being the scattering cross section corresponding to left and right circular polarization components and $I_i$ represent the intensity. The RBCs which are considered as the principle scatterers in blood lie in the Mie scattering domain. Accordingly the scattering cross section $\sigma_L$ and $\sigma_R$ are given by \cite{Born}   
\begin{equation}
\sigma_{i}=\frac{2\pi a^{2} \left|\sum_{n}(-1)^{n}{(2n+1)(a_{n,i}-b_{n,i})}\right|^{2}}{\alpha^{2}},
\end{equation}
where, 
\[a_{n,i}=\frac{\psi_{n,i}(\alpha)\psi_n^{'}(\beta)-m\psi_n(\beta)\psi'_{n,i}(\alpha)}{\xi_{n,i}(\alpha)\psi_n^{'}(\beta)-m\psi_n(\beta)\xi'_{n,i}(\alpha)}\]
and 
\[b_{n,i}=\frac{m\psi_{n,i}(\alpha)\psi_n^{'}(\beta)-\psi_n(\beta)\psi'_{n,i}(\alpha)}{m\xi_{n,i}(\alpha)\psi_n^{'}(\beta)-\psi_n(\beta)\xi'_{n,i}(\alpha)}\]
with $i=R,L$, $\alpha=k_da$, $k_d$ being the propagation constant of the radiation in ECF and $a$ is the scatterer size. 
$\beta\,(=mk_da)$ with $m$ being the relative refractive index of the scatterer (RBC in the present case). Also,  $\psi_n(\alpha)=\alpha j_n(\alpha)$ and $\xi_n(\alpha)=\alpha h^{'}_n(\alpha)$, where $j_n(\alpha)$ and $h_n(\alpha)$ represent the spherical Bessel and Hankel functions, respectively. For the present experimental setup $k_d=7.53 \mu^{-1}$m. The major light scatter with human subject is RBC. Hence the average size of the RBC is considered as $a=7\mu$m. In order to obtain the variation in refractive index as a function of glucose concentration, we have experimentally calculated $m$ (=refractive index of glucose solution/refractive index of water) 
\begin{figure}[htb]   
	\begin{center}
	\includegraphics[width=0.8\columnwidth]{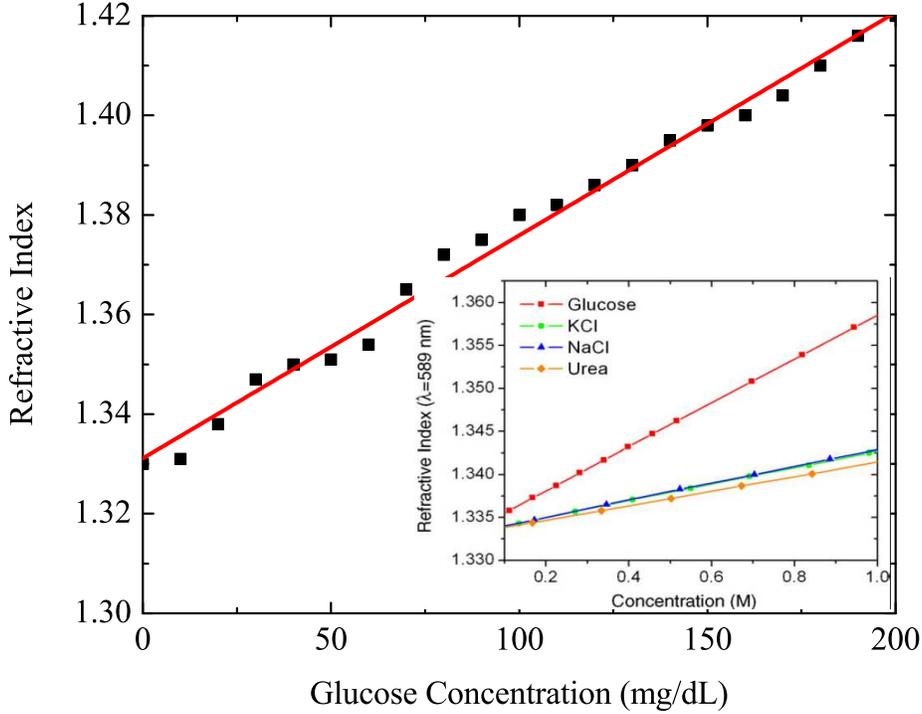}
	\caption{Linear variation in Refractive index has been with increasing glucose concentration.}
	\end{center}
\end{figure}
and exhibited in Figure 2. The inset shows the dependence of refractive index with the concentration of electrolytes (NaCl, KCl, Glucose) and Urea in water. 
It is clearly established that the contribution of glucose is dominating over others. 

Figure 3 exhibits a variation of DOCP with glucose concentration. The solid line showing a linear increase with glucose concentration is obtained using eqs. 2 and 3 and using the current experimental conditions. Data obtained with intralipid is shown (as filled dots) in the same figure for comparison.   A qualitative agreement is obtained beween the theoretical and experimental observations.
\begin{figure}[htb]   
	\begin{center}
	\includegraphics[width=0.8\columnwidth]{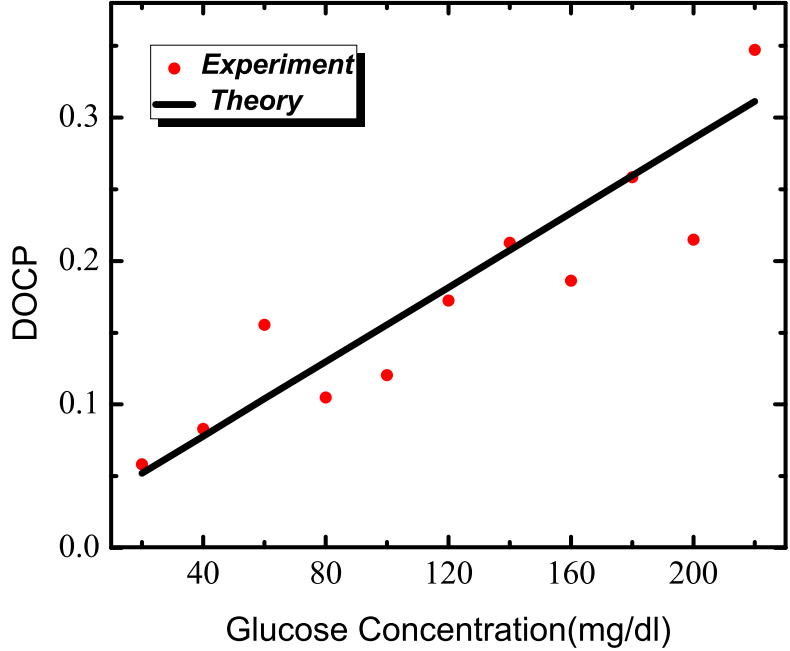}
	\caption{Linear variation in DOCP with increasing glucose concentration has observed theoretical and experimental are shown with black and red symbols,respectively.}
	\end{center}
\end{figure}

\section{Results}
Armed with the encouraging results obtained for the samples of tissue phantom, we proceed further with human subjects. In Figure 4, the degree of circular polarization obtainable from $(D_2/D_1)$ is plotted as a function of glucose concentration for four subjects.  The filled square with error-bars is obtained for voluntary subjects while the solid curve is a linear fit to the data. The slope obtained for different subjects are shown in Table 1. It is interesting to note that the ratio of DOCP and glucose concentration remains a constant for a particular subject. This fact is verified with the same subject with different conditions such as day/night, post-meal/pre-meal, etc. This fact is promising that a dedicated glucometer could be designed and fabricated for regular and hand-held usage.

\begin{figure}[htb]   
	\begin{center}
	\includegraphics[width=0.8\columnwidth]{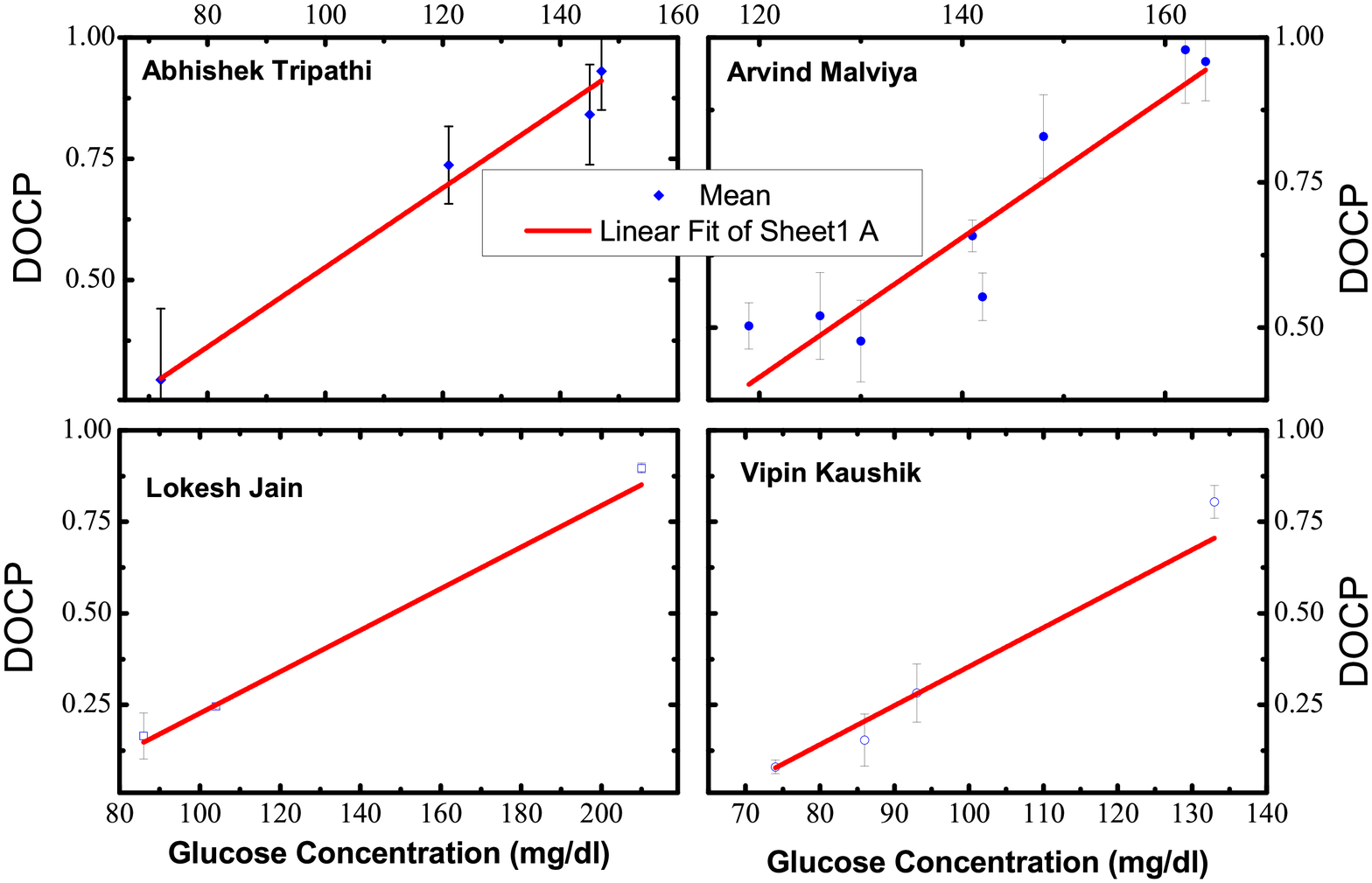}
	\caption{Simultaneous monitoring of DOCP using our setup and using commercial grade Glucometer. The results are promising.}
	\end{center}
\end{figure}

\begin{figure}[htb]   
	\begin{center}
	\includegraphics[width=0.8\columnwidth]{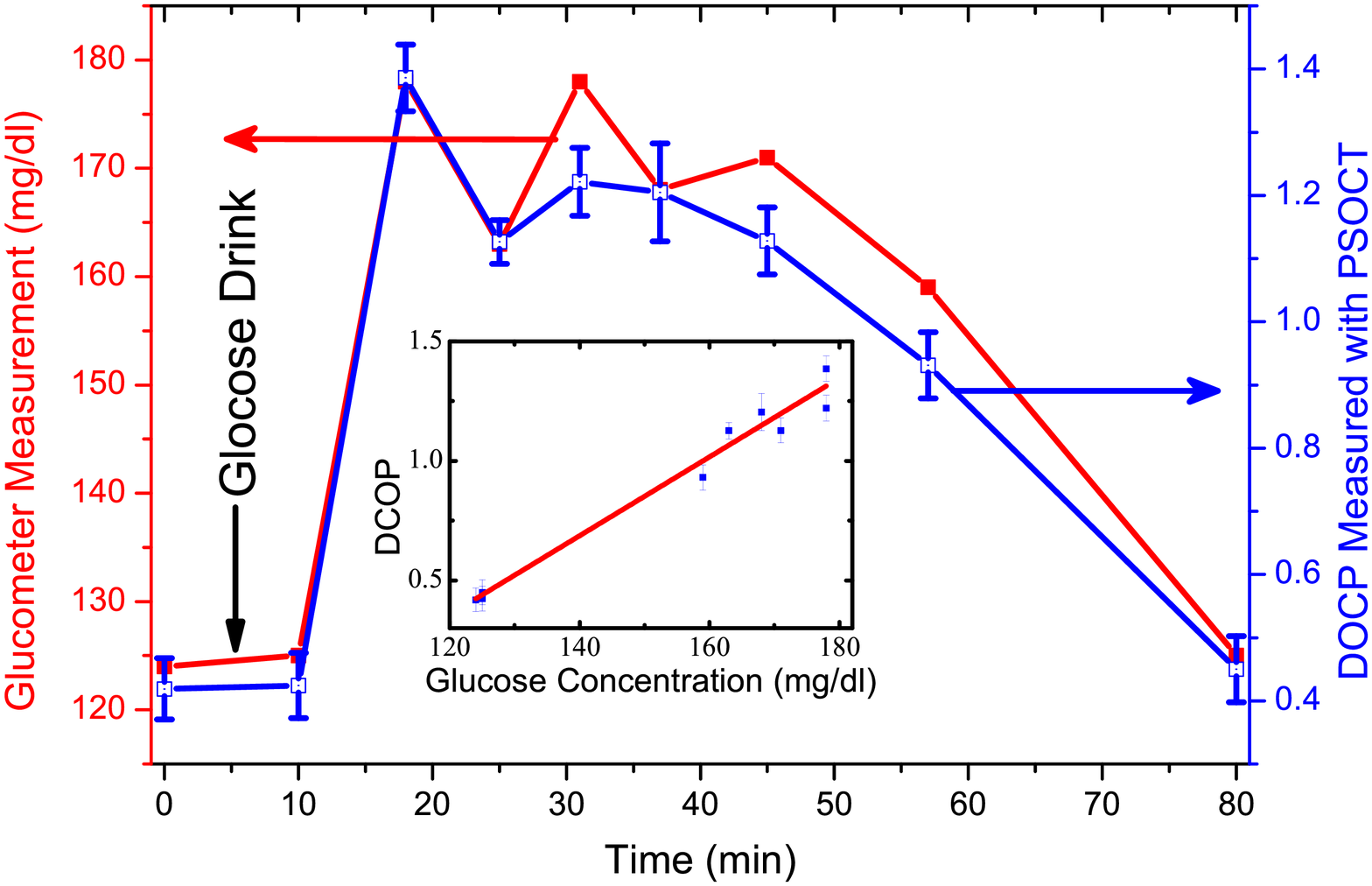}
	\caption{DOCP with increasing glucose concentration provide linearity both in theoretical and experimental which are shown with black and blue symbols,respectively.}
	\end{center}
\end{figure}

In normal subjects the blood glucose level stabilizes within 120 minutes. It is well known \cite{Varley} that the blood sugar level in human subjects initially increases after taking meal and reaches the normal value after 30-45 minutes. The normal sugar level is reached within 120 minutes \cite{Varley}. In some of the subjects, a hypoglycemic dip may also occur. After a glucose drink, we examined the subject for variation in blood glucose level in every 10-15 minutes interval. The glucose level variation as a function of time has been plotted in Figure 5. The glucometer measurements are shown as a solid square while DOCP measurements are shown by open squares with error bar. We find an excellent quantitative agreement between these curves. The corresponding data with linear fit is shown as inset. 

\section{Conclusions}
We have designed and developed PS-OCT for non-invasive glucose monitoring of blood glucose with human subjects. The results exhibited in Figures 4-5 shows that a definite linear correlation exists between the glucose concentration and the degree of circular polarization. On the other hand, we could find that the ratio of degree of circular polarization and glucose concentration is a reproducible constant value for a same person. Hence, if a hand held glucometer is developed, the slope value could be used as a signature for measuring glucose concentration for the subject under study.  

\begin{acknowledgments}
The authors thankfully acknowledged the support given by the volunters.
The authors thank Prof. P. K. Sen, SGSITS for fruitful discussions. They also acknowledge the financial support received  from UGC \& DBT, New Delhi, India and MPCOST, Bhopal, India.
\end{acknowledgments}

\begin{table}[h]
	\begin{center}
		\begin{tabular}{|c|l|c|c|c|}\hline
         Subject & Name      & Slope & RBC Size & Age \\ 
           &                 & (mg/dl)$^{-1}$      &  ($\mu$m) & Years \\ \hline \hline
         1 & A. Tripathi     & 0.03276 & 7.52 & 26\\\hline
         2 & A. Malviya      & 0.04818 & 8.10 & 31\\\hline
         3 & L. Jain         & 0.05106 & 7.90 & 30\\\hline
         4 & V. Kaushik      & 0.11702 & 8.20 & 29\\\hline
		\end{tabular}
	\end{center}
\end{table}

\end{document}